\documentclass[aps,prr,amsfonts,floatfix,superscriptaddress,longbibliography,twocolumn,footinbib]{revtex4-2}

\usepackage[utf8]{inputenc}
\usepackage{amsmath}    
\usepackage{amssymb}
\usepackage{mathbbol}
\usepackage{graphicx}   
\usepackage{verbatim}   
\usepackage{subfigure}  
\usepackage{hyperref} 
\usepackage{scalerel}
\usepackage{cancel}
\usepackage{soul}
\usepackage[normalem]{ulem}
\usepackage[dvipsnames]{xcolor}
\usepackage{siunitx}
\usepackage{physics}
\usepackage{enumerate} 
\usepackage{empheq}
\usepackage{float}

\begin{document}
\title{Instability-enhanced quantum sensing with tunable multibody interactions}

\author{Bidhi Vijaywargia}
\email[Contact author: ]{bidhi.vijaywargia@uconn.edu}
\affiliation{Department of Physics, University of Connecticut, Storrs, Connecticut 06269, USA}
\author{ Jorge Chávez-Carlos}
\affiliation{Department of Physics, Cinvestav, AP 14-740, Mexico City 07000, Mexico}
\author{Francisco Pérez-Bernal}
\affiliation{Departamento de Ciencias Integradas y Centro de Estudios Avanzados en Física, Matemáticas y Computación, Universidad de Huelva, 21071 Huelva, Spain}
\author{Lea F. Santos}
\affiliation{Department of Physics, University of Connecticut, Storrs, Connecticut 06269, USA}

\begin{abstract}
Dynamical instabilities can amplify small perturbations into measurable signals, offering a route to quantum-enhanced sensing. This mechanism was experimentally demonstrated in a collective spin system with quadratic interactions, described by a twist-and-turn Hamiltonian, where quantum evolution near an unstable point leads to exponential growth of spin fluctuations, enabling metrological gain beyond the standard quantum limit. Here, we show that a quartic extension of this Hamiltonian substantially increases the amplification. This  additional nonlinear term reshapes the phase-space structure, generating new unstable fixed points with larger local Lyapunov exponents and therefore faster signal amplification at shorter evolution times. Remarkably, even for the same instability rate (equal Lyapunov exponent), collective multibody interactions outperform the quadratic case, demonstrating that metrological gain is not determined by the Lyapunov exponent alone. We show analytically that the finite-time growth of the gain is governed also by the local phase-space curvature around the unstable point. Only for a symmetric unstable point, as in the quadratic case, is the finite-time gain described by the Lyapunov exponent alone. Our results identify local phase-space curvature as an additional resource for optimizing the speed and performance of quantum sensors.
\end{abstract}

\maketitle

\section{Introduction}

Precision measurement lies at the heart of modern science and technology, from atomic clocks and magnetometers to gravitational-wave detectors~\cite{Ludlow2015,Budker2007,Schnabel2010,Degen2017}. For independent particles, quantum fluctuations restrict measurement precision to the standard quantum limit (SQL)~\cite{Caves1981}. Overcoming this limit has motivated the development of quantum-enhanced 
metrology~\cite{Xiao1987,Sorensen2001,Giovannetti2004,Pezze2009,Pezze2018,Giovannetti2011,Toth2014}, where nonclassical correlations are used to improve sensitivity. A central approach relies on redistributing quantum fluctuations among conjugate observables, reducing uncertainty in the measured quantity at the expense of increased fluctuations in its conjugate~\cite{Wineland1992,Kitagawa1993,Ma2011}. Such squeezing-based protocols generate entanglement and enable sensitivities beyond the SQL~\cite{Pezze2018,Sorensen2001,Hu2023}.  However, their performance is constrained by the rate at which correlations can be built and by finite coherence times~\cite{Andre2002,Ma2011}.

An alternative strategy exploits dynamical instability~\cite{Li2023,Muessel2015,Manuel2023,JuliaDiaz2012,Sorelli2019}, where evolution near an unstable (hyperbolic) fixed point in phase space stretches an initially localized quantum state  along the unstable direction, enabling the amplification of weak perturbations into measurable signals. The stretching (antisqueezing) is exponentially fast in time with a growth rate set by the local Lyapunov exponent.  This mechanism has been experimentally demonstrated in a collective spin system in cavity quantum electrodynamics (QED), where quadratic collective interactions, forming a twist-and-turn Hamiltonian, yield metrological gain beyond the SQL~\cite{Li2023}.

Unstable fixed points and the associated phase-space dynamics have also been explored as a resource for generating squeezing, entanglement, and enhanced interferometric sensitivity in two-mode Bose-Hubbard Hamiltonians~\cite{JuliaDiaz2012,Sorelli2019,Strobel2014}. More recently, the role of phase-space geometry in optimizing instability-based protocols has been investigated, both by identifying the conditions under which quadratic interactions achieve locally optimal squeezing dynamics~\cite{Manuel2023} and by selecting initial spin-coherent states near hyperbolic points in quadratic two-mode models ~\cite{Shao2026}. These advances raise a fundamental question: Can dynamical instability be engineered beyond the quadratic regime to further enhance the performance of quantum sensors?

Here, we show that extending the quadratic twist-and-turn Hamiltonian with higher-order (multibody) interactions can further improve sensitivity in collective spin systems. In particular, adding a quartic spin coupling reshapes the phase-space structure, generating additional unstable points with larger local Lyapunov exponents. As a result, quantum fluctuations grow more rapidly, leading to faster amplification of weak signals. This speedup is especially important for finite-time sensing protocols, where the useful amplification time is limited in practice. A larger instability rate enables greater metrological gain in a shorter time, before decoherence degrades the signal.   

Remarkably, the multibody interaction system can outperform the quadratic model even when both have the same local Lyapunov exponent. This demonstrates that the metrological advantage is not exclusively determined by the Lyapunov exponent, but it also stems from the short-time dynamics, which depends on the local phase-space geometry around the unstable point and precedes the exponential regime. We show that two unstable points can have the same Lyapunov exponent but different curvatures along the stable and unstable directions, leading to different short-time amplifications. Therefore, quartic interactions provide an additional knob to control both the instability rate and the local phase-space structure, a freedom that is much more limited in the purely quadratic model.

Our results demonstrate that engineering dynamical instability through tailored multibody collective interactions provides a practical route to optimizing quantum sensors. The experimental platform in Ref.~\cite{Li2023} offers control over collective atomic interactions, where the same light-matter coupling used to realize quadratic terms can be extended to generate higher-order interactions via adjusted driving conditions or multiple driving frequencies~\cite{Li2023,Li2022,Colombo2022,Chen2026, Ainsworth2026}. Time-periodic (Floquet) engineering also provides a way to tailor effective interactions~\cite{Luo2025Cavity,Ma2025}. Similar capabilities arise in Kerr parametric oscillators~\cite{ZhangDykman2017, Iachello2023,Prado2025}, including superconducting circuits implementations~\cite{Sivak2019,Grimm2020,Kwon2022,Frattini2024}. Tunable multibody interactions, including three- and four-body spin interactions, have been demonstrated with trapped ions~\cite{Katz2023} and cavity platforms~\cite{Luo2025,Zhang2025}. More broadly, collective spin models have been simulated using Ising machines~\cite{Ma_Quan2026} and transmon qudits~\cite{Champion2025}. Engineering quartic interactions in these platforms enables exploration of the instability-enhanced metrological advantages identified here. 

The paper is organized as follows. Section~\ref{sec:model} introduces the model, its classical limit, and the phase diagram. Section~\ref{sec:gain} analyzes the sensing protocol and the amplification of weak perturbations. Section~\ref{sec:covariant_matrix} derives the covariance matrix result for the finite-time gain and compares the quadratic and quartic models. Section~\ref{sec:conclusion} summarizes the results and discusses experimental relevance.


\section{Collective Model with Four-body interactions}
\label{sec:model}

The model considered is motivated by cavity-QED realizations of collective spin Hamiltonians, where an ensemble of atoms couples dispersively to a single cavity mode~\cite{Li2022,Luo2025Cavity}. The two relevant internal states of the atoms are represented by collective spin operators. The cavity resonance frequency depends on their population imbalance and the intracavity field itself shifts the atomic energies. This feedback makes the effective energy nonlinear in the collective spin. A recent analytical treatment derived the complete hierarchy of these cavity-mediated nonlinearities in closed form~\cite{Ainsworth2026}. In the presence of a spin-echo sequence, odd-order contributions are canceled, leaving only even powers of the collective spin~\cite{Ainsworth2026}.

We consider the leading higher-order extension of the quadratic term, thus retaining the quadratic and quartic collective interactions. The resulting infinite-range Hamiltonian for $N=2S$ spin-$1/2$ particles is given by
\begin{equation}
\label{eq:quan_fourth_order}
\hat{H} = \Omega\, \hat{S}_z -\chi_2\, \hat{S}_x^2 - \chi_4\,\hat{S}_x^4,
\end{equation}
where $\Omega$ sets the collective rotation rate about the $z$ axis ($\Omega=2h$), while $\chi_2=2J/S$ and $\chi_4=2K/S^3$ control the quadratic and quartic nonlinearities, respectively. This Hamiltonian combines rotation about $z$ with quadratic and quartic twisting about $x$, yielding a higher-order generalization of the twist-and-turn model~\cite{Muessel2015}. Beyond quantum metrology, the same Hamiltonian has also served as a model for studying ensemble inequivalence in Refs.~\cite{Defenu2024,Daniel2026}.

The quadratic interaction, $\hat S_x^2$, realizes one-axis twisting~\cite{Kitagawa1993}, which shears the quantum fluctuations of an initially coherent spin state on the Bloch sphere, generating spin squeezing. Fluctuations are reduced along one direction and amplified along an orthogonal one, defining a squeezed quadrature that enables parameter estimation beyond the SQL~\cite{Wineland1992,Kitagawa1993,Giovannetti2004,Pezze2009,Ma2011,Pezze2018,Wineland1994}. 

Adding the linear term $ \hat S_z$ to the quadratic interaction ($\chi_2\neq 0$, $\chi_4=0$) yields the Lipkin-Meshkov-Glick (LMG) model~\cite{Lipkin1965a,Lipkin1965b,Lipkin1965c}, whose classical phase space contains a hyperbolic fixed point~\cite{Botet1983,Leyvraz2005,Santos2016}. When the system is initialized near this unstable point, small perturbations grow exponentially with a rate set by the Lyapunov exponent~\cite{Pilatowsky2020,Chavez2023}. The state is stretched along the unstable direction, producing rapid antisqueezing, and compressed along the stable direction. Combined with a time-reversal protocol, this amplification of fluctuations is converted into a measurable signal, providing an alternative route to quantum-enhanced sensitivity~\cite{Li2023}. The LMG model can be mapped to a two-mode Bose-Hubbard Hamiltonian~\cite{Steel1998,Smerzi1997,Milburn1997}, where unstable fixed points have been exploited to generate squeezing, entanglement, and enhanced interferometric sensitivity~\cite{JuliaDiaz2012, Sorelli2019}. 

In this work, we exploit the quartic collective interaction $\hat S_x^4$ as a control knob to reshape the semiclassical phase space and enhance quantum sensitivity. We show next that the quartic term  generates an additional unstable fixed point. This bifurcation changes both the local curvature and the instability rate of the unstable points, resulting in faster growth of collective spin fluctuations, as discussed in Sec.~\ref{sec:gain}.

\subsection{Phase diagram and classical limit} 

To characterize how the quartic interaction modifies the energy landscape and identify the parameter regimes where enhanced instability emerges, we analyze the classical limit of Eq.~(\ref{eq:quan_fourth_order}) and construct the corresponding phase diagram. Because the Hamiltonian in Eq.~\eqref{eq:quan_fourth_order} commutes with the total spin $\hat S^2$, the dynamics of an initial spin-coherent state is confined to the fully symmetric sector, $S = N/2$. Thus, we can describe the classical limit using a classical spin vector on the Bloch sphere. 

The classical Hamiltonian is obtained by evaluating $\hat{H}$ in a spin-coherent state 
$|\theta,\phi\rangle$ \cite{Radcliffe1971,Arecchi1972} and taking the limit $S\to\infty$ at fixed intensive
couplings (see Appendix~\ref{appen:classical}):
\begin{equation}
H_{\rm cl}(\theta,\phi)=
 2hs_z - 2Js_x^2 - 2Ks_x^4,  
\end{equation}
where $s_x=\sin\theta\cos\phi$,
$s_y=\sin\theta\sin\phi$, and 
$s_z= \cos\theta$.

For convenience, we map the Bloch sphere onto a disk using canonical coordinates 
\begin{align}
Q&=r\cos\phi ,  \nonumber \\ 
P&=-r\sin\phi ,
\end{align}
where $r=\sqrt{2(1+\cos\theta)}$. The south pole corresponds to $(Q,P)=(0,0)$ and the disk boundary $Q^2+P^2=4$ represents the north pole. 

Setting $K=0$ recovers the LMG limit. For $K\neq 0$, the quartic collective spin interaction introduces additional higher-order nonlinearities in $Q$ and $P$ into the classical Hamiltonian $H_{\rm cl}(Q,P)$. The interplay among these nonlinearities gives rise to a richer phase-space structure, which is absent for the quadratic model.

In these canonical variables, all fixed points of the Hamiltonian $H_{\rm cl}(Q,P)$ lie on the $P=0$ axis. Their number and stability can therefore be determined by the effective one-dimensional potential (see Appendix~\ref{appen:fixed_point}), 
\begin{equation}
\begin{split}
  V(Q)\equiv H_{\mathrm{cl}}(Q,0)
  =-2h+(h-2J)Q^2+\\
  \left(\frac{J}{2}-2K\right)Q^4+ KQ^6-\frac{K}{8}Q^8 .
    \label{eq:potential}
\end{split}
\end{equation}
Its shape, and consequently the fixed-point structure, is controlled by the dimensionless parameters $h/J$ and $K/J$. 

\begin{figure}[t]
\hspace*{-0.5cm}
\includegraphics[width=0.5\textwidth]{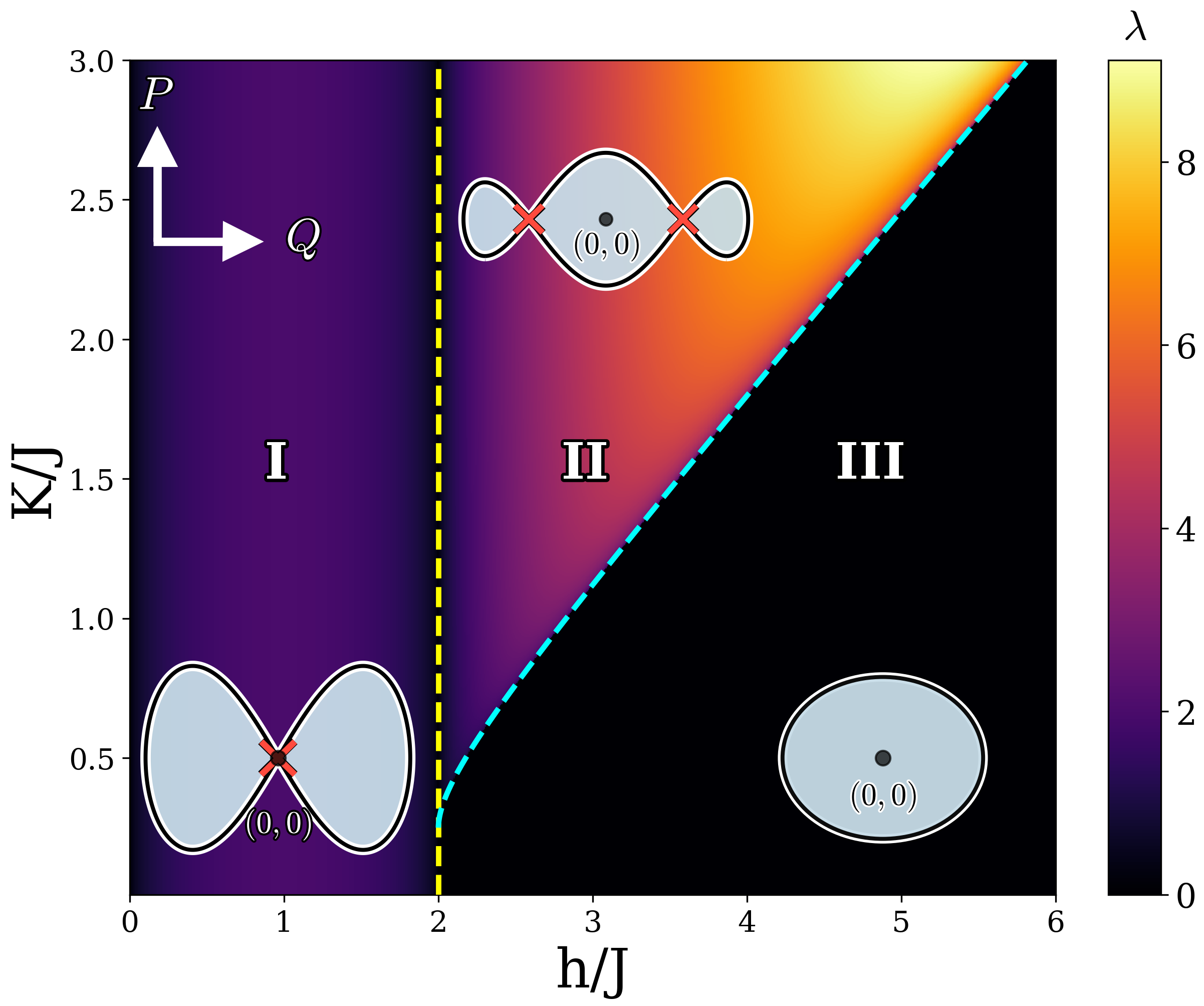}
\caption{Local Lyapunov exponent ($\lambda$) map in the ($h/J$, $K/J$) plane.  The yellow dashed line denotes $h_1/J$=2 and the cyan dashed line marks the critical value $h_2/J$ (see text). Representative phase-space structures are shown for the three regions: (I) $h/J <2$, (II) $2 <h/J < h_2/J $, and (III) $h/J > h_2/J$. Red crosses indicate hyperbolic points. In region III, there is no local $\lambda>0$.
  }
\label{fig01:PhaseDiagram}
  \end{figure}

The phase diagram is shown in Fig.~\ref{fig01:PhaseDiagram}. The yellow dashed line at 
\begin{equation}
 h_1/J=2   
\end{equation}
and the cyan dashed line at 
\begin{equation}
h_2/J = \sqrt{8(1+2K/J)^3/(27K/J)}    
\end{equation}
are obtained from the stability analysis of the classical Hamiltonian using  the one-dimensional potential in Eq.~\eqref{eq:potential} (Appendix~\ref{appen:fixed_point}). 

In region III ($h/J>h_2/J$), below the cyan line, the phase space contains a single stable fixed point at $(0,0)$, as illustrated in Fig.~\ref{fig01:PhaseDiagram}. Our focus is instead on regions I and II, where hyperbolic fixed points are present and the local Lyapunov exponent is positive. 

In region I ($h/J<2$), the classical energy landscape exhibits a double-well structure with two symmetric minima and a single hyperbolic point at the origin $(0,0)$, as shown in Fig.~\ref{fig01:PhaseDiagram}. This unstable fixed point underlies the exponential dynamics exploited for sensing in the LMG regime ($K=0$) \cite{Li2023,Manuel2023}. 

For $K/J>1/4$, the yellow dashed line at $h_1/J=2$ marks a bifurcation that defines region II ($2<h/J<h_2/J$), bounded by the two dashed lines. In this regime, the origin $(0,0)$ becomes a local stable point, while two symmetric hyperbolic points emerge at $(Q,P)=(\pm Q_{\text{hyp}}\neq0,0)$, as seen in Fig.~\ref{fig01:PhaseDiagram}. Consequently, the effective potential along $P=0$ develops a three-well structure with a central well at $Q=0$ and two outer wells separated by the hyperbolic points (see Fig.~\ref{Fig:Wells} in Appendix~\ref{appen:fixed_point}). The central and outer wells are not mutually equivalent wells.

Triple-well potentials exhibit more complex physics than double wells. Triple-well potentials arise in collective spin Hamiltonians with higher-order interactions~\cite{LohrRobles2025}, molecular and chemical dynamics~\cite{Maji2006}, nuclear two-fluid models~\cite{Gramos2016}, and Josephson amplifiers~\cite{Zorin2011, Sivak2019}. They support additional interference mechanisms and tunneling pathways~\cite{Lorch2019,Dunne2020,Prado2025}, experimentally accessible in the collective spin platforms that motivate our model. 

Notably, the triple-well structure persists even for $\chi_2=0$ ($J=0$), $\chi_4 \neq 0$ ($K\neq 0$), owing to the surviving sextic term in Eq.~\eqref{eq:potential}. Furthermore, the collective interaction $\hat S_x^4$ is not unique in generating a triple-well phase-space structure; it is simply the lowest-order even collective interaction capable of doing so. Higher even-order interactions can produce analogous phase-space topologies. For example, numerical results (not shown) indicate that adding a $\hat S_x^6$ interaction also gives rise to a triple-well structure for suitable parameters. The underlying mechanism is therefore not specific to the $S_x^4$ term, although the locations of the fixed points and the phase boundaries remain model dependent.

\subsection{Instability and Lyapunov exponent} 

The onset of dynamical instability is determined by linearizing Hamilton’s equations around the hyperbolic fixed points. This procedure yields the Lyapunov exponent $\lambda$ as a function of $Q_{\text{hyp}}$ and of the control parameters $h$, $J$, and $K$ (Appendix~\ref{appen:covariant}). 

In the double-well region, 
$\lambda=2\sqrt{h(2J-h)}$, which is maximized at $h/J=1$, leading to 
$\lambda_{\rm max}/J =2$. This value can be substantially increased in the triple-well regime, as evident in region II of Fig.~\ref{fig01:PhaseDiagram} and shown semianalytically in Appendix~\ref{appen:max_LE}. For example, at $K/J = 1.5$ and $h/J=3.26$, the quartic model gives $\lambda/J \approx 4.62$, which is more than twice the maximum value for the quadratic model. 

  \begin{figure*}[t]
      \centering
\includegraphics[width=0.98\textwidth]{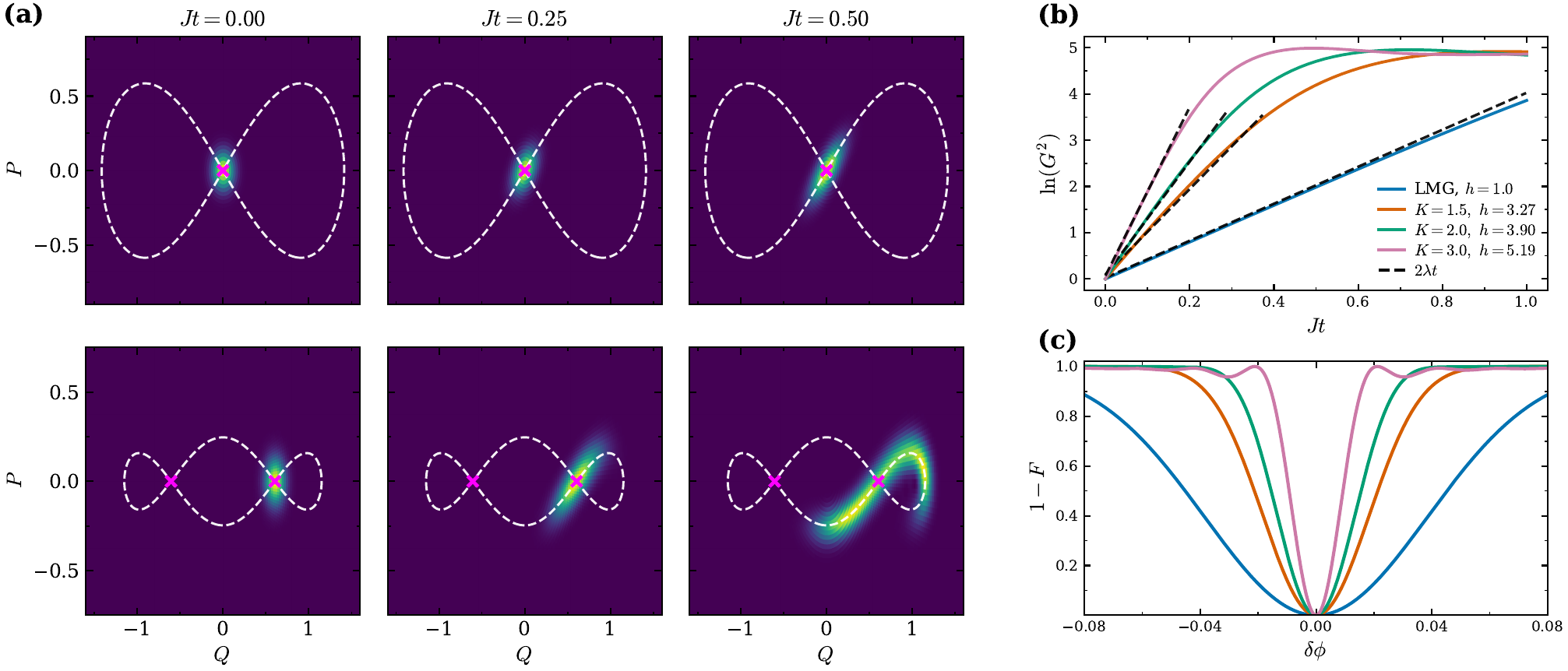}
\caption{(a) Snapshots of the Husimi functions for an initial coherent state centered at the hyperbolic point, shown for the LMG model, $K/J=0$ and $h/J=1$ (top row), and for the quartic four-body interaction model, $K/J=1.5$ and $h/J=3.265$ (bottom row), with $N=300$. The white dashed curves denote the classical separatrix, and the crosses indicate the hyperbolic points. (b) Time evolution of $\ln(G^2)$, and (c) infidelity $1-F(\delta\phi)$,  evaluated at $Jt=0.25$, as a function of the perturbation strength $\delta\phi$, for the parameters indicated in the figure; $N=500$, $J=1$. The black dashed lines in panel (b) represent the exponential growth $G^2 \propto \mathrm{e}^{2\lambda t}$ determined by the classical Lyapunov exponent.}
\label{fig02:Gain}
\end{figure*}
\section{Accelerated antisqueezing and metrological gain}
\label{sec:gain}

Having established how the quartic interaction reshapes the phase-space structure and modifies the local instability, we now examine its consequences for quantum sensing.

\subsection{Spin fluctuations, antisqueezing, and quantum Fisher information}

In collective spin systems, a spin-coherent state has isotropic quantum fluctuations in the plane perpendicular to its mean-spin direction. The variance along any transverse direction is $S/2$, which defines the SQL. Spin squeezing redistributes these fluctuations, reducing the variance along one transverse direction at the expense of increasing it along the orthogonal one~\cite{Kitagawa1993,Wineland1992}. While spin squeezing is commonly exploited to improve precision measurements, the instability-enhanced sensing protocol studied here relies instead on the rapid growth of fluctuations along the antisqueezed direction near a hyperbolic fixed point~\cite{Li2023,Manuel2023,Muessel2015,Linnemann2016}. 

The growth of these fluctuations is conventionally quantified by the antisqueezing parameter~\cite{Kitagawa1993,Ma2011,Wineland1994}, defined as follows. For a direction $\alpha$ in the plane perpendicular to the mean-spin direction at the hyperbolic fixed point, the transverse spin component is 
\[
\hat S_\alpha=\hat S_y\cos\alpha+\left(\cos\theta_{\text{hyp}} \hat S_x-\sin\theta_{\text{hyp}} \hat S_z\right)\sin\alpha,
\]
where $\theta_{\mathrm{hyp}}$ is the polar angle of the hyperbolic fixed point, with $\cos\theta_{\mathrm{hyp}} = Q_{\mathrm{hyp}}^2/2 - 1$. The antisqueezing parameter is then
\begin{equation}
\xi_+^2=\frac{\max_\alpha \mathrm{Var}(\hat S_\alpha)}{S/2}.
\end{equation}
For a spin-coherent state, $\xi_+^2=1$.
The maximization over $\alpha$ selects the direction of maximum variance $\mathrm{Var}$, corresponding to  the direction of maximum stretching (antisqueezing).  The optimal direction generally depends on time and its evolution is analyzed in Appendix~\ref{SecAppB.2}, where we also show that our conclusions persist when $\alpha$ is instead kept fixed.

The growth of the variance along the antisqueezed direction directly translates into enhanced metrological sensitivity, as quantified by the quantum Fisher information (QFI) \cite{Braunstein1994,Helstrom1976}. The QFI measures how sensitively a quantum state responds to a change in the parameter to be estimated. A larger QFI implies that smaller parameter changes can, in principle, be resolved with greater precision. For pure states, when a small parameter $\delta\phi$ is encoded through the unitary transformation $e^{-i\delta\phi \hat S_\alpha}$, corresponding to a collective rotation about the antisqueezed direction, the QFI is
\begin{equation}
   {\cal F}_Q = 4\,\mathrm{Var}(\hat S_\alpha).
\end{equation}
The variance of $\hat S_\alpha$ therefore determines how strongly the state responds to a small rotation about this direction. Since this is the same variance that characterizes the antisqueezing, the instability-induced growth of fluctuations directly enhances the QFI. Maximizing over $\alpha$ gives the maximal QFI, ${\cal F}_Q^{\max}$.

We quantify the resulting enhancement in sensitivity by the metrological gain $G^2$, defined as the maximal QFI relative to that of the initial coherent state, ${\cal F}_Q^{\mathrm{CS}}=2S$,
\begin{equation}
  G^2 =  \frac{\cal F_Q^{\max}}{\cal F_Q^\mathrm{CS}}= \frac{\max_\alpha \mathrm{Var}(\hat S_\alpha)}{S/2} = \xi_+^2 .
  \label{eq:G2}
\end{equation}
Hence, the antisqueezing parameter is precisely the metrological gain and directly
measures the metrological advantage over the SQL. We therefore use $G^2$ throughout the remainder of the paper.

The accelerated growth of the metrological sensitivity induced by the quartic interaction is illustrated in Fig.~\ref{fig02:Gain}(a), which shows snapshots of the Husimi function (phase-space distribution) for an initial coherent state centered at the hyperbolic point. The top row corresponds to the LMG model and the bottom row shows the quartic interaction case. While the state spreads along the unstable direction in both cases, the expansion is markedly faster for the quartic model, consistent with its larger local Lyapunov exponent $\lambda$.

At the hyperbolic point, the dynamical instability stretches the quantum state along the unstable (antisqueezed) direction. This stretching grows as $e^{\lambda t}$, so that the corresponding variance grows as $\mathrm{Var}(\hat S_\alpha)\propto e^{2\lambda t}$. From Eq.~(\ref{eq:G2}), this exponential growth directly translates into the metrological gain,
\begin{equation}
G^2 \propto e^{2\lambda t}.
\end{equation}

Figure~\ref{fig02:Gain}(b) compares the time dependence of $\ln G^2$ for the LMG and  quartic models at different parameter values. Multibody interactions substantially accelerate the growth of the gain. While the LMG model has a local Lyapunov exponent $\lambda = 2$, the quartic model exhibits considerably steeper slopes. This enhancement persists even when $\alpha$ is fixed rather than optimized along the direction of maximal variance [see Fig.~\ref{figSM:alpha_gain}(b) in Appendix~\ref{SecAppB.2}].

The comparison in Fig.~\ref{fig02:Gain}(b) is performed at the same quadratic interaction strength $J$, showing that the quartic interaction increases the instability rate and therefore reaches a given metrological gain in a shorter time. This is important for realistic settings, where finite coherence times limit the time window over which amplification can be effectively exploited. A comparison at fixed Lyapunov exponent, which isolates the role of the local phase-space geometry, is presented in Sec.~\ref{sec:covariant_matrix}.

\subsection{Time-reversal protocol}

To convert the growth into a measurable signal, we employ a time-reversal protocol~\cite{Li2023,Davis2016,Macri2016}. The system first evolves forward, generating antisqueezing. A perturbation is then applied in the plane orthogonal to the mean spin, followed by  backward evolution, 
\begin{equation}
\label{eq:satin}
|\psi_t(\delta \phi)\rangle
= \hat U_t(\delta \phi) |\psi_0\rangle  = e^{+i\hat H t} e^{-i\delta \phi \hat S_\alpha} e^{-i\hat H t} |\psi_0\rangle .
\end{equation}
Because the forward evolution strongly stretches the state along the unstable direction, the backward evolution converts the small rotation into a large collective spin signal,
\begin{equation}
\langle \hat S_\alpha \rangle = S \sin(G\,\delta \phi),
\end{equation}
where the amplification factor $G$ is precisely the gain introduced in Eq.~(\ref{eq:G2}) \cite{Davis2016,Nolan2017}. In the unstable regime, this amplification grows exponentially, $G^2 \propto e^{2\lambda t}$, enabling metrological gain. 

To further characterize the sensitivity of the time-reversal protocol, we compute the fidelity 
\begin{equation}
   F(\delta \phi) = |\langle \psi_0 | \psi_t(\delta \phi) \rangle|^2 . 
\end{equation}
A rapid decay of $F(\delta \phi)$ indicates that the protocol amplifies small rotations into highly distinguishable states.

In the limit of small $\delta \phi$, the fidelity is related to the metrological gain as 
\begin{equation}
 F(\delta \phi) \simeq 1 - \frac{S}{2} G^2 \delta \phi^2 . 
\end{equation}
Thus, the curvature of the fidelity and the metrological gain are equivalent measures of sensitivity in the small-rotation regime.

Figure~\ref{fig02:Gain}(c) shows the infidelity $1-F(\delta \phi)$ as a function of the perturbation strength $\delta \phi$. While all cases exhibit the expected quadratic behavior at small $\delta\phi$, the presence of multibody interactions leads to a significantly steeper growth with $\delta \phi$, consistent with larger metrological gain at this fixed evolution time.

\section{Finite-time gain beyond Lyapunov exponent}
\label{sec:covariant_matrix}

\begin{figure}[t]
      \centering
\includegraphics[width=0.4\textwidth]{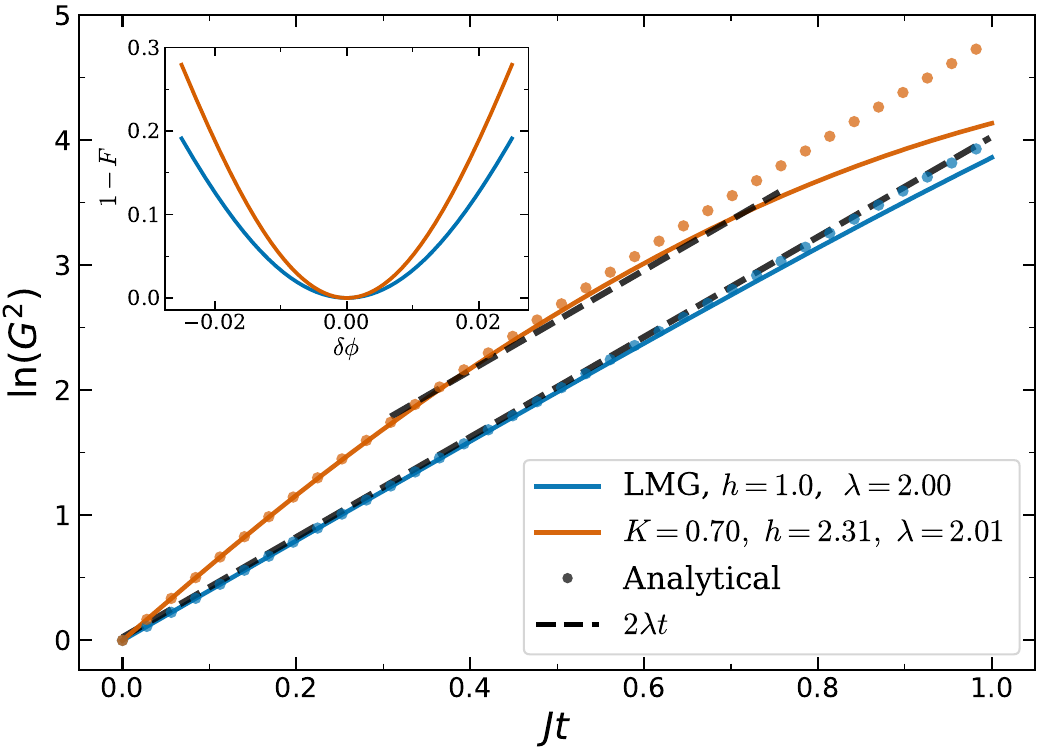}
\caption{Time evolution of $\ln(G^2)$ for the LMG and quartic models with parameters chosen such that both have nearly the same Lyapunov exponent $\lambda \sim 2$. Solid lines: numerical results; black dashed lines: Lyapunov  growth; circles: analytical result in Eq.~(\ref{eq:gain_analytical}); $N=500$, $J=1$. Inset: infidelity, $1-F(\delta\phi)$, for both models evaluated at $Jt=0.25$.  
}
      \label{fig03}
  \end{figure}

We showed in the previous section that the quartic interaction can generate larger local Lyapunov exponents than the quadratic model, leading to faster amplification of quantum fluctuations. In this section, we address a different question: Is the finite-time metrological gain determined only by the Lyapunov exponent? Interestingly, the answer is no. Figure~\ref{fig03} compares the LMG and quartic models for parameters chosen close to the bifurcation line (the yellow dashed line in Fig.~\ref{fig01:PhaseDiagram}), for which both systems have nearly the same local Lyapunov exponent, $\lambda/J \simeq 2$.  Despite having the same instability rate, the quartic model exhibits a noticeably larger metrological gain. As we explain next, this demonstrates that sensing depends not only on the Lyapunov exponent, but also on the local geometry of the phase space around the hyperbolic fixed point.

To understand the enhanced gain, we analyze the local dynamics around the hyperbolic fixed point using the covariance-matrix formalism~\cite{Ma2011}, which tracks how an initially localized distribution spreads and deforms in phase space (see Appendix~\ref{appen:covariant} for details). Expanding the classical Hamiltonian to quadratic order in the displacements $\delta Q = Q- Q_{\text{hyp}}$ and $\delta P = P$ from the unstable point gives
\begin{equation}
\label{eq:local_expansion}
H_{\rm local}\approx
\mu\,\delta Q^2+\nu\,\delta P^2,
\end{equation}
where 
\begin{equation}
\begin{split}
\mu &= h - 2J + 3JQ_{\text{hyp}}^2 - 12KQ_{\text{hyp}}^2 + 15KQ_{\text{hyp}}^4 - \tfrac{7}{2}KQ_{\text{hyp}}^6, \\
\nu  &= h + \tfrac{1}{2}JQ_{\text{hyp}}^2 + KQ_{\text{hyp}}^4 - \tfrac{1}{4}KQ_{\text{hyp}}^6,
\end{split}
\end{equation}
which implies that for the LMG model, where $Q_{\text{hyp}}=0$ and $K=0$, we have $\mu = h - 2J$ and $\nu  = h$.
The corresponding linearized dynamics is fully characterized by the covariance matrix of the phase-space fluctuations. Solving its time evolution yields the analytical expression
\begin{equation}
\label{eq:gain_analytical}
\ln(G^2)
=
2\sinh^{-1}
\!\left[
\frac{\kappa}{\lambda}
\sinh(\lambda t)
\right],
\end{equation}
where
\[
\lambda=2\sqrt{-\mu\nu}
\]
is the local Lyapunov exponent and
\[
\kappa=
\left|
\frac{\nu}{\tau^2}
-
\mu\tau^2
\right|,
\qquad
\tau^2
=
1-\frac{Q_{\rm hyp}^2}{4}.
\]
The factor $\tau$ accounts for the mapping between the Bloch sphere and the canonical phase-space coordinates [see Eq.~(\ref{eq:spin_qp})]. 

For a hyperbolic fixed point, $\mu<0$ and $\nu>0$, so that
$$ \kappa= 2 \left( \frac{ |\mu|\tau^2+\frac{\nu}{\tau^2} }{2} \right) , \qquad \lambda=2\sqrt{\left(|\mu|\tau^2\right) \left(\frac{\nu}{\tau^2}\right)}. $$
Thus, $\kappa$ and $\lambda$ are twice the arithmetic and the geometric means, respectively, of the two local curvatures
$|\mu|\tau^{2}$ and $\nu/\tau^{2}$. Since the arithmetic mean is always greater than or equal to the geometric mean, $\kappa\mathrel{\geqslant}\lambda$ for
any hyperbolic fixed point, with equality only in the symmetric case,
$\nu=|\mu|\tau^{4}$.

Equation~(\ref{eq:gain_analytical}) reveals that the gain depends on two independent local quantities. The Lyapunov exponent $\lambda$ determines the exponential growth, whereas $\kappa$ characterizes the local phase-space curvature around the hyperbolic fixed point. In the short-time limit,
\[
\sinh(\lambda t)\ll\frac{\lambda}{\kappa},
\]
Eq.~(\ref{eq:gain_analytical}) reduces to
\begin{equation}
 \ln(G^2)\approx2\kappa t,  
 \label{Eq:2kappat}
\end{equation}
showing that the initial logarithmic growth rate is governed by $\kappa$. At longer times, the gain crosses over to the Lyapunov growth with slope $2\lambda$. Therefore, deviations from the Lyapunov growth arise at short times when $\kappa>\lambda$.

Figure~\ref{fig03} confirms these predictions. The quartic and LMG models are chosen to have nearly identical Lyapunov exponents, $\lambda/J\simeq 2$, yet the quartic model exhibits a larger gain. For the LMG model, the maximum Lyapunov exponent $\lambda_{\rm max}/J=2$ is reached at $J=h$, where the hyperbolic point is symmetric and $\kappa=\lambda$. Equation~(\ref{eq:gain_analytical}) then reduces to the purely exponential growth, $G^2=e^{2\lambda t}$. In contrast, the quartic model can realize the same Lyapunov exponent with $\kappa>\lambda$. It follows from Eqs.~(\ref{eq:gain_analytical}) and (\ref{Eq:2kappat}) that a faster initial growth of the gain is achieved, with logarithmic slope $2\kappa$, before crossing over to the Lyapunov growth with slope $2\lambda$. Consequently, the quartic interaction provides a finite-time enhancement beyond that expected from the Lyapunov exponent alone. See also Fig.~\ref{fig:analytical} in Appendix~\ref{SecAppB.1}.

The inset of Fig.~\ref{fig03} shows the infidelity $1-F$ for both models at fixed time $Jt=0.25$. While both models exhibit the expected quadratic behavior at small $\delta \phi$, the quartic model shows a significantly steeper growth, consistent with its enhanced metrological gain.

The analytical result also reveals an important difference between the LMG and quartic models. While the LMG model exhibits an intrinsic trade-off between the instability rate and the finite-time enhancement associated with the local phase-space curvature, the quartic interaction removes this constraint.

For the LMG model, the hyperbolic fixed point always lies at $Q_{\rm hyp}=0$, so that $\kappa_{\rm LMG}=2J$, independent of $h$, whereas $\lambda_{\rm LMG}=2\sqrt{h(2J-h)}\mathrel{\leqslant}2J$. As explained above, the maximum Lyapunov exponent is reached at $h/J=1$, where $\lambda_{\rm LMG}=\kappa_{\rm LMG}=2J$. Away from this point, the LMG model can achieve $\kappa/\lambda>1$, but only at the expense of reducing the Lyapunov exponent. In other words, the LMG model exhibits a trade-off between the instability rate and the finite-time enhancement associated with the local phase-space curvature. 

The quartic interaction removes this constraint. Owing to the additional hyperbolic fixed points created by the quartic term, one can simultaneously realize $\lambda>2J$ and $\kappa/\lambda>1$. Consequently, the quartic model combines two independent mechanisms for enhancing metrological gain: a larger Lyapunov exponent, which accelerates the asymptotic amplification of fluctuations, and a curvature-induced finite-time enhancement governed by $\kappa/\lambda$.

Linearized semiclassical and Gaussian descriptions of  two-mode Bose-Hubbard Hamiltonians have been used to describe the short-time growth of spin fluctuations, squeezing, and entanglement~\cite{JuliaDiaz2012,Sorelli2019}. Here, we employ a similar local expansion around the hyperbolic fixed point for a different purpose: to identify the local phase-space curvature as an independent resource governing metrological gain. The quartic interaction makes this resource tunable through the emergence of additional hyperbolic fixed points.

\section{Discussion and Conclusion}
\label{sec:conclusion}

We have shown that collective multibody interactions provide a route for enhancing instability-based quantum sensing. Extending the quadratic twist-and-turn Hamiltonian with a quartic collective interaction reshapes the semiclassical phase space through a bifurcation from a double- to a triple-well structure, generating new hyperbolic fixed points with larger local Lyapunov exponents. As a result, quantum fluctuations are amplified more rapidly, enabling larger metrological gain at shorter evolution times.

Beyond increasing the instability rate, we demonstrated that the metrological gain is not determined by the Lyapunov exponent alone. Using a covariance-matrix description of the local dynamics, we derived an analytical expression for the gain in terms of two independent local quantities: the Lyapunov exponent, which governs the asymptotic exponential amplification, and a curvature parameter, which controls the short-time dynamics. While these two quantities coincide for the symmetric hyperbolic point of the model with quadratic interaction only, the quartic interaction lifts this constraint, allowing systems with the same Lyapunov exponent to exhibit different finite-time gains. The quartic model can therefore exploit two complementary mechanisms simultaneously: faster amplification through a larger Lyapunov exponent and an additional finite-time enhancement arising from the local phase-space curvature.

More broadly, our results establish local phase-space geometry as a resource for optimizing instability-enhanced quantum metrology. The quartic collective spin model provides a minimal and experimentally motivated example in which both the instability rate and the local curvature can be engineered by tuning the parameters of the system. This suggests that similar mechanisms may be explored in other nonlinear platforms possessing unstable fixed points, such as Kerr parametric oscillators.

The present work focuses on unitary dynamics. An important direction for future research is to investigate the robustness of the finite-time enhancement in the presence of decoherence, imperfect time reversal, and other experimental imperfections, as well as to explore whether higher-order collective interactions can provide further metrological advantages.

\begin{acknowledgments}
This work was supported by the Research Corporation Cottrell SEED Award No. CS-SEED-2025-003. F.P.-B. thanks the funding received through Grant No. PID2022-136228NB-C21 funded by MICIU/AEI/10.13039/501100011033 and, as appropriate, by “ERDF A way of making Europe, by ERDF/EU,” by the European Union, or by the European Union NextGenerationEU/PRTR. Computing resources supporting this work were
partly provided by the CEAFMC and Universidad de Huelva High Performance Computer located in the Campus Universitario “El Carmen” and funded by FEDER/MINECO Project No. UNHU-15CE-2848. J.C.-C. has received a fellowship (Grant No. 493338) from Secretaría de Ciencia, Humanidades, Tecnología e Innovación (SECIHTI). L.F.S. acknowledges support from the U.S. Department of Energy under Grant No. DE-SC0026657.
\end{acknowledgments}

\section*{Data availability}
The data that support the findings of this paper can be generated with the code available on Zenodo~\cite{Vijaywargia2026code}. 


\appendix

\section{Classical Hamiltonian and Fixed Points}
\label{appen:classical}
Here, we derive the classical Hamiltonian corresponding to Eq.~(\ref{eq:quan_fourth_order}) and analyze its phase-space structure, fixed points, and their stability. 

\subsection{Classical Hamiltonian}
The quantum Hamiltonian in Eq.~(\ref{eq:quan_fourth_order}) is in a rotated basis relative to Ref.~\cite{Li2023}. Its classical counterpart is obtained from its expectation value in a spin-coherent state, $|\theta,\phi\rangle$~\cite{Radcliffe1971,Arecchi1972}, in the limit $S\to\infty$,
  \begin{equation}
  \label{eq:classical_fourth_def}
  H_{\mathrm{cl}}
  =\lim_{S\rightarrow\infty}\frac{\langle\theta,\phi|\hat H|\theta ,\phi \rangle}{S}
  =2(hs_z-Js_x^2-Ks_x^4),
  \end{equation}
where
\[
s_x = \sin\theta\cos\phi,
 \qquad s_y = \sin\theta\sin\phi, \qquad 
s_z = \cos\theta
\]
are the classical spin components on the Bloch sphere and 
\begin{equation}
  |\theta,\phi\rangle
  =\frac{1}{(1+|\zeta|^2)^S}\sum_{m=-S}^{S}\zeta^{S-m}
  \binom{2S}{S+m}^{1/2}|S,m\rangle,
  \end{equation}
with $\zeta=\tan(\theta/2)e^{i\phi}$.

To cast the dynamics in canonical phase-space form, we map the Bloch sphere onto a disk using 
\begin{equation}
\label{eq:disk_sphere}
Q=r\cos\phi,  \,\,\, P=-r\sin\phi, \,\, r=\sqrt{2(1+\cos\theta)},
\end{equation}
for which $\{Q,P\}=1$. The south pole is mapped to $(Q,P)=(0,0)$, while the north pole lies on the boundary $Q^2+P^2=4$, so the dynamics is restricted to the disk $Q^2+P^2\mathrel{\leqslant} 4$. In these variables, the spin components read
\begin{equation}
s_x = Q\sqrt{1-\dfrac{r^2}{4}},\,\,
s_y = -P\sqrt{1-\dfrac{r^2}{4}},\,\,
s_z = \dfrac{r^2}{2}-1 ,
\label{eq:spin_qp}
\end{equation}
and the Hamiltonian becomes  \begin{equation}
  \begin{split}
  H_{\mathrm{cl}}(Q,P)=&-2h+h(Q^2+P^2)-2JQ^2+\frac{J}{2}Q^2P^2+\frac{J}{2}Q^4 \\
  &-2KQ^4+KQ^4P^2-\frac{1}{8}KQ^4P^4
  \\
  &+KQ^6-\frac{1}{4}KQ^6P^2-\frac{1}{8}KQ^8 .
  \label{Eq:HclQuartic}
  \end{split}
  \end{equation}
  Setting $K=0$ recovers the LMG limit, whereas $K\neq 0$ generates higher-order terms in both $Q$ and $P$. 

\subsection{Fixed points}  
\label{appen:fixed_point}
The fixed points satisfy Hamilton's equations 
\[
\dot Q=\partial_P H_{\mathrm{cl}}, \qquad \dot P=-\partial_Q H_{\mathrm{cl}}.
\]
All fixed points lie on the $P=0$ axis, since 
\[
\dot{Q} = P[\,2h + JQ^2 + 2KQ^4(1-r^2/4)\,]
\]
and the bracketed factor is positive inside the disk for $h,J,K>0$. As a result, the fixed points and their stability can be determined by the effective one-dimensional potential 
  \begin{equation}
  \begin{split}
  V(Q)\equiv H_{\mathrm{cl}}(Q,0)
  =-2h+(h-2J)Q^2+\\ \left(\frac{J}{2}-2K\right)Q^4
  +KQ^6-\frac{K}{8}Q^8 .
  \end{split}
  \end{equation}
The fixed points are determined by $V'(Q) =0$, while their stability is determined by the Hessian, which in this case is determined by $V''(Q)$ evaluated at those fixed points. The equation $V'(Q) = 0$ always admits the trivial solution, $Q =0$, implying that $(Q,P)=(0,0)$ is a fixed point of the Hamiltonian for all parameter values. 
\begin{figure*}
    \centering
    \includegraphics[width=0.99\textwidth]{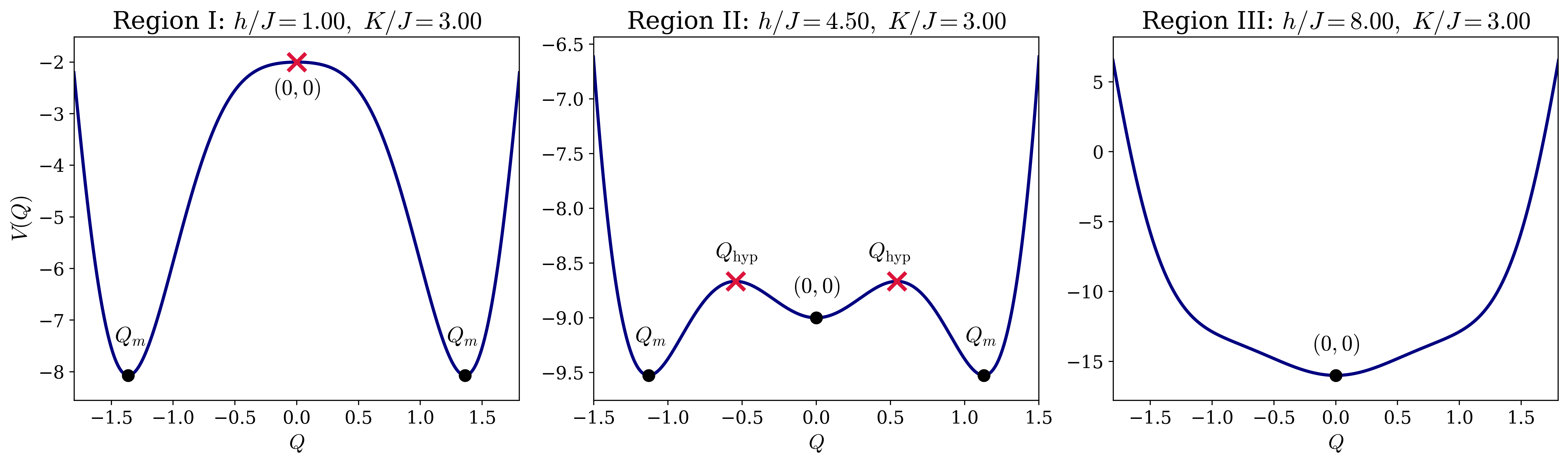}
    \caption{Effective potentials $V(Q)=H_{\rm cl}(Q,0)$ in the three regions of the phase diagram. In region I ($h/J<2$), the potential has a double-well structure with two minima at $
  \pm Q_m$ (black dots) and a hyperbolic fixed point at the origin (red cross). In region II ($2<h/J<h_2/J$), the potential develops a triple-well structure with a central minimum at $(0,0)$, outer
  minima at $\pm Q_m$, and two hyperbolic fixed points at $\pm Q_{\mathrm{hyp}}$. In region III ($h/J>h_2/J$), only a single minimum at the origin remains and no hyperbolic fixed points are present.}
  \label{Fig:Wells}
\end{figure*}
For the LMG model ($K=0$), the effective potential reduces to 
$$V_{\mathrm{LMG}} = -2h+(h-2J)Q^2+\frac{J}{2}Q^4.$$
The fixed points are therefore 
 \begin{equation}
  (Q,P)=(0,0),
  \qquad
  (Q,P)=\left(\pm\sqrt{{2-h/J}},\,0\right),
  \end{equation}
where the nonzero pair is positive only for $h/J<2$. Thus, for $h/J < 2$, the phase space contains three fixed points and the effective potential has a double-well structure. In this region, the origin is hyperbolic, while the two symmetric nonzero fixed points correspond to the minima of the potential. On the other hand, for $h/J> 2$, the origin is the only fixed point and corresponds to the minimum of the potential. The line $h/J=2$ marks the change of stability of the origin and corresponds to the LMG model ground-state phase transition. 

For $K \neq 0$, the nonzero fixed points satisfy a cubic equation, which in terms of the dimensionless parameters $h/J$ and $K/J$ can be written as 
\begin{equation}
  x^3-6x^2-\frac{2(1-4K/J)}{K/J}\,x-\frac{2(h/J-2)}{K/J}=0,
  \label{eq:cubic_dimless}
\end{equation}
where $x=Q^2$. Only positive roots satisfying $0<x\mathrel{\leqslant}4$ correspond to physical fixed points. This cubic equation admits different solutions depending on the parameters. We distinguish different regions based on the number and the nature of the fixed points (see Fig.~\ref{fig01:PhaseDiagram}). 

An important threshold occurs at $K/J=1/4$. For $K<J/4$, the phase diagram contains only two physical regions, namely, regions I and III in Fig.~\ref{fig01:PhaseDiagram}, while the intermediate region II is absent. To understand this, consider the line $h/J=2$, where Eq.~\eqref{eq:cubic_dimless} factorizes as
  \begin{equation}
  x\left[x^2-6x-\frac{2(1-4K/J)}{K/J}\right]=0.
  \end{equation}
The solutions at $h/J=2$ show that an additional nonzero physical branch emerges only when the quartic coupling exceeds the threshold $K/J=1/4$. Consequently, for $K/J<1/4$, the phase diagram contains only regions I and III: for $h/J<2$, there is one nonzero physical pair of fixed points and the origin (region I), whereas for $h/J>2$, the only fixed point is the origin (region III). Thus, for $K=0$ and, more generally, for $K/J<1/4$, the phase diagram contains only regions I and III, separated by the yellow dashed line at $h_1/J=2$ in Fig.~\ref{fig01:PhaseDiagram}.

For $K\neq0$ and $K/J>1/4$, the intermediate region II appears. To determine its boundary, it is convenient to rewrite Eq.~\eqref{eq:cubic_dimless} by introducing the
shifted variable $x=z+2$, which brings the cubic equation to the depressed form
  \[
  z^3+cz+d=0,
  \]
  with $c=-\frac{2(1+2K/J)}{K/J}$ and $d=-\frac{2h/J}{K/J}$. The discriminant of this cubic equation is
  \[
  D=\frac{32(1+2K/J)^3-108(h/J)^2(K/J)}{(K/J)^3}.
  \]
The transition line is obtained from the condition $D=0$, which yields
  \begin{equation}
  \label{eq:ht_dimless}
  \frac{h_2}{J}
  =
  \sqrt{\frac{8\left(1+2K/J\right)^3}{27\,K/J}}.
  \end{equation}
Accordingly, for $K\neq0$ and $K/J>1/4$, the phase diagram is partitioned into three distinct regions separated by the lines $h_1/J=2$ and $h_2/J$, shown, respectively, by the yellow and cyan dashed lines in Fig.~\ref{fig01:PhaseDiagram}. 

The fixed-point structure in each region is determined by the number of positive physical roots of Eq.~\eqref{eq:cubic_dimless} and the stability of the corresponding fixed points is determined by the $V''(Q)$. Each region is then characterized as follows.

(1) Region I: $0<h/J<h_{1}/J$. 
In this regime, Eq.~\eqref{eq:cubic_dimless} has a single positive physical root, denoted as $x_m=Q_m^2$. This gives one symmetry-related pair of nonzero fixed points $(\pm Q_m,0)$. The origin $(0,0)$ is hyperbolic, while the nonzero pair corresponds to the minima of the effective potential. The phase portrait therefore has a double-well structure (left panel of Fig.~\ref{Fig:Wells}).

(2) Region II: $2<h/J<h_2/J$ and $K/J>1/4$. 
In this regime, Eq.~\eqref{eq:cubic_dimless} admits two positive physical roots, which generate two symmetry-related pairs of nonzero fixed points. One pair corresponds to hyperbolic fixed points, $(\pm Q_{\mathrm{hyp}},0)$, while $(\pm Q_m,0)$ corresponds to global or local minima. Together with the fixed point at the origin, this yields five physical fixed points in total. The location of the fixed points, $Q_{\mathrm{hyp}}$ and $Q_m$, depends on the parameters $h/J$ and $K/J$ and therefore varies across the phase diagram. The corresponding effective potential has a triple-well structure (middle panel of Fig.~\ref{Fig:Wells}). In this region, the origin $(0,0)$ is no longer a hyperbolic fixed point, as it was in region I, but
instead corresponds to a global or local minimum. This additional structure is generated by the sixth-order term of the potential in Eq.~(\ref{eq:potential}), while the eighth-order term controls the relative depth of the outer wells. The coexistence of multiple competing minima signals a first-order quantum phase transition. 

(3) Region III: $h/J>h_2/J$.
In this regime, Eq.~\eqref{eq:cubic_dimless} has no positive physical roots. The only remaining fixed point is the origin $(0,0)$, which corresponds to
the unique minimum of the effective potential (right panel of Fig.~\ref{Fig:Wells}). No hyperbolic fixed points exist in this region.

Notably, setting $J=0$ does not remove the triple-well structure, since the effective potential still contains the sixth- and eighth-order terms induced by the quartic
interaction. This suggests that, even in the absence of the quadratic nonlinearity $S_x^2$, the interplay between the $S_x^4$ term and the field term $S_z$ can produce rich quantum behavior. 

\subsection{ Fixed points on the Bloch sphere}

The fixed points obtained in the $(Q, P)$ phase space can be mapped back onto the Bloch sphere using the inverse of the transformation in Eq.~\eqref{eq:disk_sphere}. Since all fixed points lie on the $P = 0$ axis, one has $\phi = 0$ for $Q > 0$ and $\phi = \pi$ for $Q < 0$, with the polar angle given by
  \[
  \cos\theta=\frac{Q^2}{2}-1.
  \]
Each symmetry-related pair $(\pm Q_*, 0)$ maps to a pair of symmetric point on the Bloch sphere, while the origin $(0, 0)$ corresponds to the south pole $(\theta = \pi)$. In the following Appendixes, we work primarily in the canonical $(Q, P)$ variables, as they provide a more natural framework for analyzing the linearized dynamics and covariance evolution near the hyperbolic fixed points.

\section{Antisqueezing and Covariance matrix}
\label{appen:covariant}

As illustrated in the main text, growth of variance near the hyperbolic fixed point is central to the metrological enhancement. In this Appendix, we connect this growth to the linearized flow and fluctuations in the phase space through the covariance matrix formalism {\cite{Ma2011,Weedbrook2012,Simon1994}}.

\subsection{Antisqueezing under linearized flow}
\label{SecAppB.1}

In this Appendix, we derive an analytic expression for $\xi_+^2$ (equivalently, $G^2)$ for short times using the classical covariance matrix formalism. Our starting point is a coherent state centered at the hyperbolic fixed point, $(Q_{\text{hyp}},0)$. To characterize its early-time evolution, we locally expand the Hamiltonian about this point in  terms of small displacements $\delta Q = Q- Q_{\text{hyp}}$ and $\delta P = P$. 

Under this expansion, the classical Hamiltonian for the quartic model [Eq.~(\ref{Eq:HclQuartic})] takes the form 
\begin{equation}
\label{eq:Hlocal_cl}
H_{cl}
= H_0
+ \mu\,\delta Q^2
+ \nu\,\delta P^2
+ \gamma\,\delta Q^3
+ \eta\,\delta Q\,\delta P^2
+ \mathcal{O}(\delta^4),
\end{equation}
where $H_0 = H_{\mathrm{cl}}(Q_{\text{hyp}}, 0)$ is the classical  trajectory at the hyperbolic point, and the expansion coefficients are 
\begin{equation}
\begin{split}
\mu &= h - 2J + 3JQ_{\text{hyp}}^2 - 12KQ_{\text{hyp}}^2 + 15KQ_{\text{hyp}}^4 - \tfrac{7}{2}KQ_{\text{hyp}}^6, \\
\nu  &= h + \tfrac{1}{2}JQ_{\text{hyp}}^2 + KQ_{\text{hyp}}^4 - \tfrac{1}{4}KQ_{\text{hyp}}^6,\\
\gamma &= Q_{\text{hyp}}\bigl(2J - 8K + 20KQ_{\text{hyp}}^2 - 7KQ_{\text{hyp}}^4\bigr), \\
\eta   &= Q_{\text{hyp}}\bigl(J + 4KQ_{\text{hyp}}^2 - \tfrac{3}{2}KQ_{\text{hyp}}^4\bigr).
\end{split}
\label{eq:local_coeff}
\end{equation}

At early times, $\delta Q$ and $\delta P$ are very small and the cubic and higher-order terms are negligible. The dynamics is then governed by the quadratic Hamiltonian, 
\begin{equation}
    H_{local} = \mu\,\delta Q^2
+ \nu\,\delta P^2 + \text{constants}.
\end{equation}
The resulting equations of motion are linear and can be written as \[
\frac{d}{dt}
\begin{pmatrix}
\delta Q \\
\delta P
\end{pmatrix}
= A
\begin{pmatrix}
\delta Q \\
\delta P
\end{pmatrix},
\]
where 
$$A= \begin{pmatrix}
0 & 2\nu \\
-2\mu & 0
\end{pmatrix}$$ is  the Jacobian matrix. The largest eigenvalue of this matrix gives the positive Lyapunov exponent, $\lambda = 2\sqrt{-\mu \nu}$, which for LMG case ($Q_{\mathrm{hyp}}=0, \, \mu =h-2J, \,\nu = h$) reduces to $\lambda\rm_{LMG} = 2\sqrt{h(2J-h)}$. The condition $\mu <0$ and $\nu >0$ ensures $\lambda>0$, which holds for a hyperbolic fixed point. The solution to the equations of motion is 
\[
\begin{pmatrix}
\delta Q(t) \\
\delta P(t)
\end{pmatrix}
= M(t)
\begin{pmatrix}
\delta Q(0) \\
\delta P(0)
\end{pmatrix},
\qquad
M(t) = e^{At}.
\]
Note that $A^2 = -4\mu\nu\mathbb{1} = \lambda^2\mathbb{1}$; thus, 
\begin{equation}
\begin{split}
& M(t) = \cosh(\lambda t)\mathbb{1} + \frac{\sinh(\lambda t)}{\lambda}A \quad = \\
& \hspace{25mm} \begin{pmatrix} \cosh(\lambda t) & \dfrac{2\nu}{\lambda}\sinh(\lambda t)\\ 
 \dfrac{-2\mu}{\lambda}\sinh(\lambda t) & \cosh(\lambda t) \end{pmatrix}.
\end{split}
\nonumber
\end{equation}
The covariance matrix is defined as a $2 \times 2$ symmetric matrix, 
\begin{equation}
    \Gamma_\mathrm{C} = \begin{pmatrix} \mathrm{Var}(\delta Q) & \mathrm{Cov}(\delta Q, \delta P) \\ \mathrm{Cov}(\delta Q, \delta P) & \mathrm{Var}(\delta P) \end{pmatrix},
\end{equation}
where $\mathrm{Cov}(\delta Q,\delta P)
=\langle \delta Q\,\delta P \rangle - \langle \delta Q \rangle \langle \delta P \rangle.$ The diagonal elements give the variance of the distribution along $Q$ and $P$ directions, whereas the off-diagonal elements characterize the correlation between the fluctuations in $Q$ and $P$. 
At time $t$, the covariance matrix evolves as
\begin{equation}
\label{eq:covariant_evolution}
  \Gamma_{\mathrm{C}}(t) = M(t)\,\Gamma_{\mathrm{C}}(0)\,M^{\mathrm{T}}(t).
\end{equation}
$\Gamma_{\mathrm{C}}(t)$ encodes how an initial distribution in $(Q,P)$, localized at the hyperbolic fixed point, deforms over time under the linearized flow. Its eigenvalues give the variances along the maximally stretched (unstable) and maximally compressed (stable) directions, while the corresponding eigenvectors identify these directions.

The classical covariance matrix is the phase-space counterpart of the quantum covariance matrix built from the transverse spin fluctuations, 
 
\begin{equation}
  \Gamma_{\mathrm{Q}} =
  \begin{pmatrix}
  \mathrm{Var}(\hat S_{\vec e_1}) & \mathrm{Cov}(\hat S_{\vec e_1},\hat S_{\vec e_2}) \\
  \mathrm{Cov}(\hat S_{\vec e_1},\hat S_{\vec e_2}) & \mathrm{Var}(\hat S_{\vec e_2})
  \end{pmatrix},
\end{equation}
  where   $
  \mathrm{Cov}(\hat S_{\vec e_1},\hat S_{\vec e_2})
  =\frac{1}{2}\Bigl\langle\bigl[\hat S_{\vec e_1}, \hat S_{\vec e_2}\bigr]_+\Bigr\rangle-\langle \hat S_{\vec e_1} \rangle \langle \hat S_{\vec e_2} \rangle $.
Here, $\hat S_{\vec e_1}$ and $\hat S_{\vec e_2}$ denote collective spin components transverse to the mean-spin direction. The antisqueezing is then given by the largest eigenvalue of $\Gamma_\mathrm{Q}$ \cite{Ma2011} normalized to $S/2$, the initial variance of the coherent state. 
\begin{figure}[h]
\centering
\includegraphics[width=0.46\textwidth]{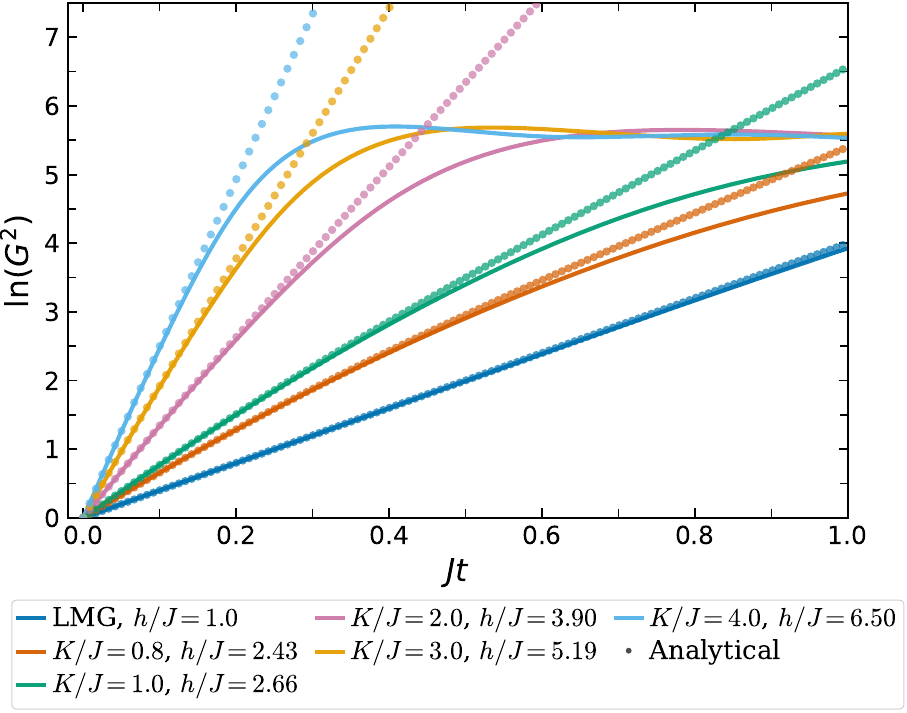}
\caption{Time evolution of $\ln G^2(=\xi_\mathrm{+}^2)$ for the LMG and quartic models with parameters shown in the legends. Solid lines represent numerical results, while the circles are the analytical results in Eq.~(\ref{eq:log_antisqz_analytic}); $N=1000$, $J=1$.}  
\label{fig:analytical}
\end{figure}
We define $\xi_\mathrm{C}$ as the classical counterpart of the quantum antisqueezing and it is given by the largest eigenvalue of $\Gamma_\mathrm{C}$ normalized to the initial distribution. A spin-coherent state centered at the hyperbolic fixed
point is isotropic in the local tangent frame of the Bloch sphere. In the canonical \((Q,P)\) coordinates, this becomes
\begin{equation}
    \Gamma_C(0)=\mathrm{diag}(\tau^2,1/\tau^2),
    \qquad
    \tau^2=1-\frac{Q_{\rm hyp}^2}{4}.
\end{equation}
The factor \(\tau\) follows from the mapping between the Bloch-sphere variables and \((Q,P)\) [Eq.~\eqref{eq:spin_qp}]. At the south pole, \(Q_{\rm hyp}=0\), one has \(\tau=1\), and the coherent state is circular in \((Q,P)\).

\begin{figure*}
    \includegraphics[width=0.95\textwidth]{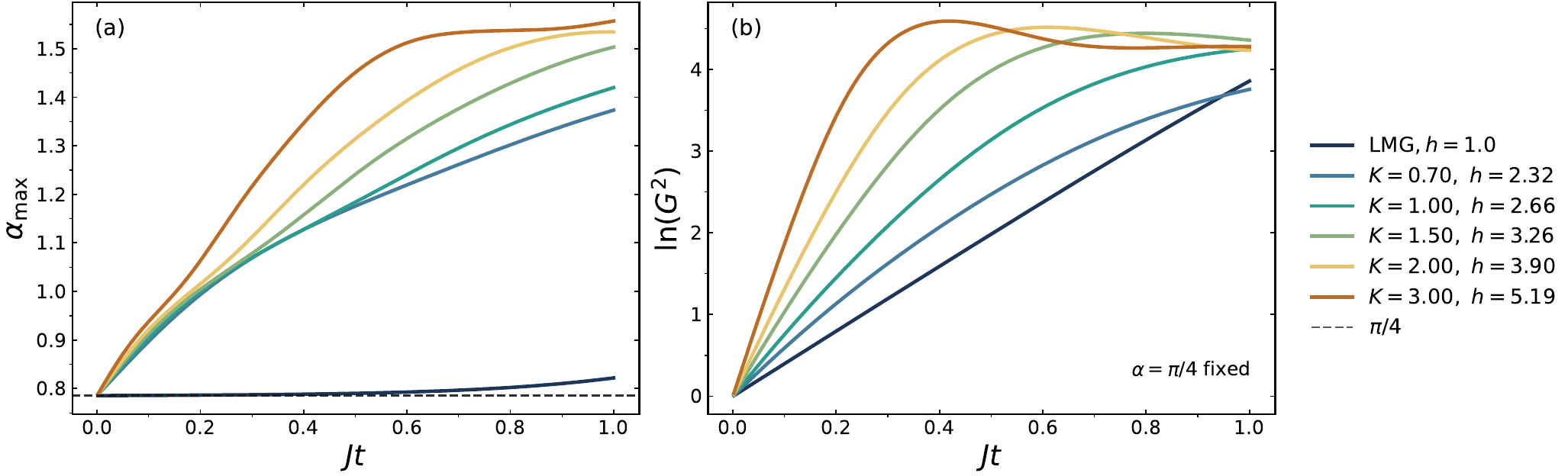}
    \caption{
  (a) Time evolution of the optimal antisqueezing angle $\alpha_{\max}$ obtained by maximizing the antisqueezing over $\alpha$ at each instant of time. The black dashed horizontal line marks $\alpha=\pi/4$.
  (b) Corresponding growth of gain, shown as $\ln G^2$, evaluated at the fixed angle $\alpha=\pi/4$ for the same set of parameter $(K/J,h/J)$ values as in panel (a). The cases for the quartic model ($K>0$) exhibit substantially
  stronger antisqueezing growth than the LMG model, showing that the enhancement persists even without time-dependent optimization of $\alpha$. Here, $N=500$.
  \label{figSM:alpha_gain}}
  
\end{figure*}

To find how this initial distribution stretches with time, we solve for the largest eigenvalue of $\Gamma_{\mathrm{C}}(t)$ [Eq.~\eqref{eq:covariant_evolution}] normalized to the initial distribution. This gives a closed-form formula of antisqueezing as   
\begin{equation}
\label{eq:antisqz_analytical} 
\xi_\mathrm{C}^2(t)=\left(\sqrt{1+\rho(t)}+\sqrt{\rho(t)}\right)^2,
\,\,
\rho(t)=\frac{\kappa^2}{\lambda^2}\sinh^2(\lambda t),
\end{equation}
where $\kappa = |\nu/\tau^2-\mu\tau^2|$ = $|\tilde \nu -\tilde \mu|$.

Further, 
\begin{equation}
\ln(\xi_C^2) = 2\,\sinh^{-1}\!\left[(\kappa/\lambda)\sinh(\lambda t)\right].
\label{eq:log_antisqz_analytic}
\end{equation}

In the semiclassical limit $S \rightarrow \infty$, the local canonical coordinates $(\delta Q, \delta P)$ provide the classical description of the transverse spin fluctuations. Therefore, $\xi_{+}^2 \approx \xi_\mathrm{C}^2$ before the higher-order terms in the expansion of the Hamiltonian become relevant. This is illustrated in Fig.~\ref{fig:analytical}. 

For the LMG model, the hyperbolic point sits at $Q_\mathrm{hyp}=0$, so $\tau^2=1$. At $h/J=1$ (where $\lambda$ is maximized) $\nu=-\mu$ and therefore $\kappa=\lambda$, and Eq.~(\ref{eq:antisqz_analytical}) simplifies to
\begin{equation}
  \xi_C^2(t)=e^{2\lambda t},
  \end{equation}
corresponding to purely exponential growth. In contrast, away from special symmetric points, the bifurcated hyperbolic points of the quartic model have unequal local curvatures, \(|\tilde \mu|\neq \tilde \nu\). Then, \(\kappa>\lambda\), and the finite-time gain deviates from the pure Lyapunov prediction. Equation~\eqref{eq:antisqz_analytical} shows that the growth of $\xi_+^2$ ($G^2$) in the quartic model is not determined by the Lyapunov exponent $\lambda$ alone, but also depends on the local structure of the phase-space flow near the hyperbolic fixed point. This can lead to a stronger enhancement than expected from $\lambda$ by itself.

\subsection{Time dependence of optimal direction}
\label{SecAppB.2}

The direction of maximal variance (antisqueezed direction) is given by the eigenvector of $\Gamma_\mathrm{C}(t)$ associated with its largest eigenvalue. At time $t=0$, the normalized covariance matrix is proportional to identity, so there is no preferred direction. For short times, in the local tangent-frame variables,
this is governed by the symmetric part of the linearized flow, \(A+A^{\rm T}\). For the hyperbolic Hamiltonian considered here, this matrix has eigenvectors \((1,\pm1)/\sqrt{2}\). The growing direction is therefore initially aligned with \(\alpha=\pi/4\).

For the LMG model at $h/J=1$, one has $\nu=-\mu$ (and $\tau^2=1$). In this symmetric case, $\Gamma_{\rm C}(t)$ has equal diagonal entries, and the antisqueezed direction remains aligned with $\alpha=\pi/4$ as long as the local quadratic approximation is valid. In contrast, for the quartic model, $|\tilde \mu| \neq \tilde \nu$ and the asymmetry between the diagonal entries of $\Gamma_{\rm C}(t)$ causes the antisqueezed direction to rotate away from $\pi/4$, as shown in Fig.~\ref{figSM:alpha_gain}(a).

Even without tracking this time-dependent rotation, evaluating the gain at the fixed angle $\alpha=\pi/4$ already yields substantial growth in the quartic model. For the parameter values shown in Fig.~\ref{figSM:alpha_gain}(b), this fixed angle gain exceeds the corresponding LMG result. Thus, even though  $\alpha=\pi/4$ is no longer the instantaneous direction of maximal variance in the quartic model, it still has significant overlap with the unstable direction and captures much of the finite-time amplification.

\subsection{Maximum Lyapunov exponent}
\label{appen:max_LE}

\begin{figure}
\includegraphics[width=0.45\textwidth]{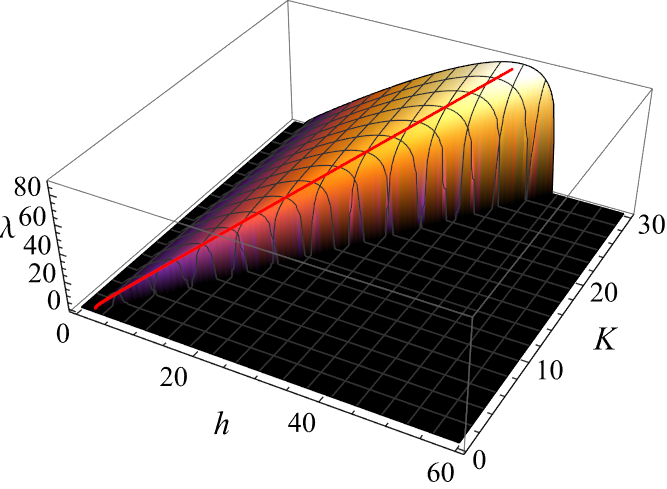}
\caption{Local Lyapunov exponent as a function of $h$ and $K$ for $J=1$. The red line corresponds to Eq.~(\ref{Eq:MaxLE}) and represents the maximum Lyapunov exponent along the surface of $\lambda$. }
\label{figSM:MaxLyap}
\end{figure}

As evident from the bright region in Fig.~\ref{fig01:PhaseDiagram}, the strongest classical instability in region II occurs approximately along a straight line in the $(h/J,K/J)$ plane. Numerically, for $J=1$, this optimal line is well described by
\begin{equation}
  K_{\mathrm{opt}} \approx 0.7547\,h - 0.8691.
  \label{Eq:MaxLE}
\end{equation}
This relation provides a useful guide for identifying parameter values that maximize the classical Lyapunov exponent, as shown in Fig.~\ref{figSM:MaxLyap}.

\section{Local expansion around the hyperbolic point}

In this Appendix, we extend the local expansion of the quartic Hamiltonian beyond the quadratic order and compare the leading nonlinear corrections in the quartic and LMG models. 

\subsection{Classical expansion}

As discussed in Appendix~\ref{appen:covariant}, the local expansion of the quartic Hamiltonian near the hyperbolic point takes the form  
\begin{equation}
H_{cl}
= H_0
+ \mu\,\delta Q^2
+ \nu\,\delta P^2
+ \gamma\,\delta Q^3
+ \eta\,\delta Q\,\delta P^2
+ \mathcal{O}(\delta^4),
\label{eq:local_classical_cubic}
\end{equation}
where the coefficients are given in Eq.~(\ref{eq:local_coeff}). For the LMG model ($K = 0$, $Q_{\text{hyp}} = 0$), all cubic terms vanish identically. More generally, even for $K \neq 0$ the cubic terms vanish whenever $Q_{\mathrm{hyp}}=0$ , which is the case for $h < 2J$. In such situations, the leading correction to the quadratic approximation arises only at fourth order. 
However, this does not hold for the generic quartic case with $Q_{\mathrm{hyp}} \neq 0$, where the cubic terms are present and provide the leading nonlinear correction to the local dynamics.

\subsection{Quantum bosonic form}

To obtain the quantum counterpart of the local Hamiltonian near the hyperbolic point, we express the fluctuations in terms of bosonic operators,
  \[
  \delta Q=\frac{a+a^\dagger}{\sqrt{2}},
  \qquad
  \delta P=\frac{a-a^\dagger}{i\sqrt{2}},
  \]
with $[a,a^\dagger]=1$ and $\hat n=a^\dagger a$. Upon quantization and symmetrization of the classical Hamiltonian in Eq.~\eqref{eq:local_classical_cubic}, the local quantum
Hamiltonian takes the form
\begin{equation}
\begin{split}
\hat{H}_{\mathrm{local}} = H_0 + \frac{\mu - \nu}{2}\bigl(a^2 + a^{\dagger 2}\bigr)
+ (\mu + \nu)\,\hat{n}
+ \frac{\mu + \nu}{2} \\
+\frac{\gamma - \eta}{2\sqrt{2}}\bigl(a^{\dagger 3} + a^3\bigr)
+ \frac{3\gamma + \eta}{2\sqrt{2}}
  \bigl(a^{\dagger 2}a + a^\dagger a^2 + a^\dagger + a\bigr).
\end{split}
\label{eq:Hlocal_quantum}
\end{equation}
The cubic corrections in Eq.~\eqref{eq:Hlocal_quantum} provide the leading nonlinear deviations from the quadratic dynamics once the wave packet has spread beyond the strictly local regime. In the quartic model, these cubic terms are present when \(Q_{\rm hyp}\neq0\), reflecting the fact that the phase-space landscape is locally asymmetric on the two sides of the hyperbolic point. They cause the phase-space distribution to distort unevenly as it spreads along the separatrix. 

For LMG, setting $K = 0$ and $Q_{\text{hyp}} = 0$ gives $\mu = h - 2J$, $\nu= h$, and $\gamma = \eta = 0$, so that
{\predisplaypenalty=0
\begin{equation}
\label{eq:H_LMG}
\hat{H}_{\mathrm{local}}^{\,\mathrm{LMG}}
\approx H_0
- J\bigl(a^2 + a^{\dagger 2}\bigr)
+ 2(h - J)\,\hat{n}
+ (h - J),
\end{equation}}
recovering a pure squeezing Hamiltonian (two-axis countertwisting) ~\cite{Kitagawa1993,Manuel2023,Caves1981} at the critical value $h/J = 1$.

It is important to note that, at early times and sufficiently close to the hyperbolic fixed point, the dynamics of both models is governed by the quadratic Hamiltonian. The cubic corrections become relevant once the wave packet spreads along the separatrix. 


%



\end{document}